\documentclass[prl,twocolumn,superscriptaddress,showpacs,nofootinbib]{revtex4}
\usepackage{graphicx}

\usepackage{bm}
\usepackage{amssymb}
\usepackage{epstopdf}
\usepackage{amsmath}
\usepackage{amstext}
\usepackage{latexsym}
\usepackage{hyperref}
\usepackage{pstricks}
\usepackage{ulem}

\newcommand{\ket}[1]{\left\vert#1\right\rangle}

\begin{document}

\title{Probing magnetic order in ultracold lattice gases} 
\author{G. De Chiara}
\affiliation{F\'isica Te\`orica: Informaci\'o i Processos Qu\`antics, Universitat Aut\`{o}noma de Barcelona, E-08193 Bellaterra, Spain}
\author{O. Romero-Isart}
\affiliation{Max-Planck-Institut f\"ur Quantenoptik,
Hans-Kopfermann-Strasse 1,
D-85748, Garching, Germany.}
\author{A. Sanpera}
\affiliation{ICREA, Instituci\`o Catalana de Recerca i Estudis Avan\c{c}ats, E08011 Barcelona}
\affiliation{F\'isica Te\`orica: Informaci\'o i Processos Qu\`antics, Universitat Aut\`{o}noma de Barcelona, E-08193 Bellaterra, Spain}

\begin{abstract}
A forthcoming challenge in ultracold lattice gases is the simulation of quantum magnetism. That involves both, the preparation of the lattice atomic gas in the
desired spin state as well as the probing of the state.  Here we demonstrate how a probing scheme, based on atom-light interfaces, gives access to the {\it order parameters} of non trivial quantum magnetic phases,  allowing to characterize univocally strongly correlated magnetic systems produced in ultracold gases. The method, which furthermore is non demolishing, yields spatially resolved spin correlations and can be applied to bosons or fermions. As a proof of principle, we apply the method to detect the complete phase diagram displayed by a chain of (rotationally invariant) spin-1 bosons.
\end{abstract}

\maketitle

The possibility of using ultracold gases as quantum simulators, i.e. an experimental system that mimics
models of condensed matter, high energy physics or quantum chemistry, was boosted by the seminal realization of Bose-Einstein condensate in alkaline atoms \cite{BEC}. Since then, the field has progressed enormously, both theoretically and experimentally, moving from the weakly-correlated regime towards the strongly-correlated one \cite{reviewBloch,reviewAnna}. This has been accomplished mainly by the use of optical lattices and Feshbach resonances. A forthcoming step is the simulation of quantum magnetism --spin ordering-- in ultracold lattice gases. Recent experiments, showing super-exchange interactions in two nearby sites or novel methods to lower the temperatures enough to achieve this goal have already been reported \cite{TrotzkyAndKetterle}.\\
\indent One of the major difficulties in simulating magnetic systems is not only the preparation but also the faithful detection of the many-body state. A 
tomographic reconstruction of the state is normally unrealistic due to the number of particles involved in the system. 
However, in quantum magnetism, ordering is usually reflected by a set of {\it order parameters}. An order parameter is a local operator $\hat{A}$
with a non vanishing expectation value, $\langle \varphi |\hat{A}|\varphi \rangle\neq 0$, that accounts for some broken symmetry embedded in the ground state $\ket{\varphi}$. Landau introduced the order parameter concept as a way to quantify the dramatic transformation of matter at a (classical) phase transition and since then it has become a cornerstone concept in physics. Notice that there exist also quantum phases like e.g. topological, whose symmetries are not local but global and, therefore, cannot be described by local order parameters. In these systems, strong quantum fluctuations  prevents the formation of local ordering.\\
\begin{figure}[t]
\begin{center}
\includegraphics[scale=0.7]{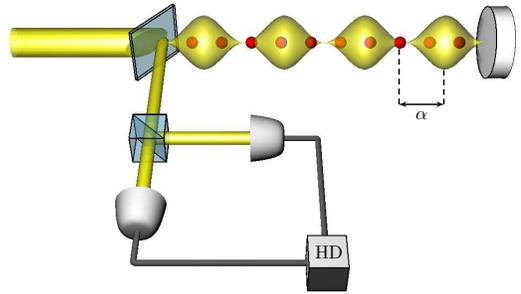}
\caption{(Color online) Schematic detection setup, atoms placed in an optical lattice of periodicity $d/2$ (not shown) are illuminated by a laser beam in a standing wave configuration (yellow region) shifted by $\alpha$ from the optical lattice configuration. The output light is redirected through a polarimeter which measures its polarization through a homodyne detection (HD).}
\label{fig:setup}
\end{center}
\end{figure}
\indent While the total magnetization will be the appropriate order parameter for a ferromagnetic state  
($\langle {\cal{M}}_z\rangle\neq0$; ${\cal{M}}_z=\sum_i S_{zi}$, where $S_{zi}$ is the operator of the $i$th spin), the characterization of spin ordering in systems with zero magnetization requires different order parameters. Sometimes order parameters are not observables (e.g. the superfluid order parameter) and cannot be directly measured. Notice that any local (or global) order parameter could be extracted from the knowledge of all order spin-spin correlations since the later
is equivalent to knowing the ground state. Detection methods are, however, restricted to lower moments. For instance, standard detection 
methods like noise-correlations \cite{Altman2004,Bloch2005} detect density-density correlations which can be linked 
to spin-spin correlations, but are not suitable for extracting higher order spin correlations. As proposed in \cite{Hulet2010}, Bragg scattering of light gives rise to the spin structure factor, useful to detect N\'{e}el antiferromagnetic states, but will fail to detect 
other states that have zero magnetization and do not break time-reversal symmetry.  An alternative technique for imaging spin states in optical lattices has been put forward \cite{Burnett}. Single lattice detectors \cite{markusandbloch} should be able to resolve magnetic ordering although they are presently limited to the type of atoms they can detect and constitute a destructive technique.

\begin{figure}[!t]
\includegraphics[scale=0.7]{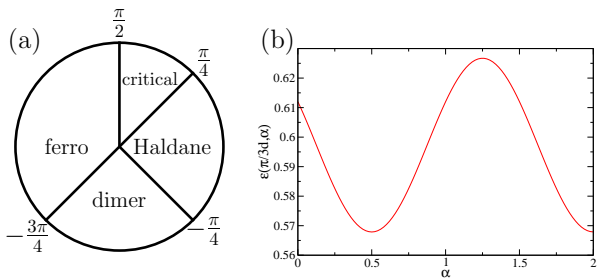}
\caption{(Color online) (a) Phase diagram of the model of Eq.\eqref{eq:HBB}. (b) The quantity $\varepsilon(\pi/3d,\alpha)$ (in units of $\kappa^2$) for $\theta=0.3 \pi$  for $L=132$ as a function of $\alpha$.}
\label{fig:2}
\end{figure}

A novel tool to detect magnetic correlations in lattice gases, based on quantum polarization spectroscopy (QPS), can overcome these limitations. The method, reported in Ref.~\cite{Eckert2008} and depicted schematically in Fig.~\ref{fig:setup}, takes advantage of  the Faraday effect occurring when off-resonant light interacts with an atomic system (either bosonic or fermionic) with internal spin degrees of freedom. A polarized light pulse  sent to a spin-polarized atomic sample couples its polarization to the atomic spin. As a result of the interaction, the light polarization rotates proportionally to the sample magnetization. Indeed, QPS is based in an effective dipole interaction Hamiltonian where an adiabatic elimination of the excited states has been performed \cite{Julsgaard}:
\begin{equation}
H_{\text{eff}}= -\kappa s_3 J_z^{\text{eff}}.
\end{equation}
The coupling constant $\kappa$ depends on the optical depth of the atomic sample and on the spontaneous emission rate. A light pulse propagating along the z-axis and strongly polarized in the x-direction is conveniently represented by Stokes operators  
$
s_1=\frac{1}{2}(a^\dagger_x a_x-a^\dagger_y a_y),
s_2=\frac{1}{2}(a^\dagger_y a_x+a^\dagger_x a_y),
s_3=\frac{1}{2i}(a^\dagger_y a_x-a^\dagger_x a_y)
$
where $a_x$ ($a_y$) is the annihilation operator of a photon with $x$ ($y$) polarization.
The atomic magnetic moments are included in the effective angular momentum:  $J^{\text{eff}}_z =1/\sqrt{L} \sum_n c_n S_{zn}$ where, 
$L$ is the total number of spins . For strongly localized atomic Wannier functions, the coefficients $c_n$, related to the standing wave intensity profile, can be approximated by: $c_n=2\cos^2[k_Pd(n-\alpha)]$ with $k_P$ being the wave number of the probing laser, $d$ the lattice spacing, and $\alpha$ the dimensionless shift between the probing laser and the lattice. 
The operator  $J^{\text{eff}}_z$ is mapped onto the light quadrature $X=\int dt s_2/\sqrt{N_{ph}}$ where we have assumed the incoming pulse to be strongly polarized along the x- direction,  $\langle (\int dt s_1)\rangle=N_{ph} \gg 1$.  The result of the mapping is \cite{Julsgaard}:
\begin{equation}
X_{out} = X_{in} - \kappa J_z^{\text{eff}}
\end{equation}
where $X_{out}$ ($X_{in}$) is the quadrature of the outgoing (incoming) light. A homodyne measure of $X_{out}$ gives access not only 
to the effective angular momentum $J_z^{\text{eff}}$  but also to the variance (as well as higher moments), demonstrating the inherent quantum character of the probe. Let us focus on the variance $\varepsilon(k_P,\alpha)$ to prove how QPS can be linked to different order parameters:
\begin{equation}
\label{eq:eps}
\varepsilon(k_P,\alpha)\equiv(\Delta J_z^{\text{eff}})^2=\frac 1L \sum_{nm}c_m c_n\mathcal G_z(m,n).
\end{equation}
where $\mathcal G_z(m,n)\equiv\langle S_{zm} S_{zn}\rangle -\langle S_{zm}\rangle\langle S_{zn}\rangle $ is the two-point spin correlation function. 
As noticed in \cite{Roscilde2009}, the  magnetic structure factor, defined as $S(q) = 1/L \sum_{mn} \exp[i qd(m-n)]\langle S_{zm}S_{zn}\rangle$,
can be connected to the variance $\varepsilon(k_P,\alpha)$ as the average over $\alpha$:
\begin{eqnarray}
\bar\varepsilon(k_P) \equiv\int d\alpha\varepsilon(k_P,\alpha)=
\frac{1}{2}S(2k_P).
\end{eqnarray}
Although the magnetic structure factor contains the two-body correlation functions, it does not provides the same information as order parameters since by averaging over the phase shift $\alpha$ one loses some information on the correlations.  In this work instead we assume an accurate control on the shift $\alpha$ and we introduce the quantity:
\begin{equation}
\label{eq:deltaeps}
\Delta\varepsilon(k_P,\alpha_1,\alpha_2)\equiv \varepsilon(k_P,\alpha_1)-\varepsilon(k_P,\alpha_2)
\end{equation}
which is the difference of the signal with fixed wavevector $k_P$ at two different phase shifts. By appropriately choosing the parameters $k_P, \alpha_1, \alpha_2$,  this quantity is linked to local order parameters able to detect not only ferro and antiferromagnetic (N\'{e}el) order, but also more exotic spin ordering. This makes QPS a valuable tool for quantum simulators. We support our claim analyzing a one dimensional spin-1 chain, a model characterized by different magnetic orderings and that can be realized with ultracold atoms in optical lattices.\\
\indent Spin-1 atoms confined in a deep optical lattice are well described by the Bose-Hubbard Hamiltonian \cite{Imambekov}.  For unit filling and for sufficiently small tunneling the system is in a Mott insulator state with one atom per site. Virtual tunneling of the atoms between neighboring sites gives rise to an effective magnetic interaction described by the bilinear-biquadratic Hamiltonian \cite{Imambekov}:
\begin{equation}
\label{eq:HBB}
H_{BB} =\sum_i \cos(\theta) \bm{S}_i\cdot\bm{S}_{i+1}+ \sin(\theta) (\bm{S}_i\cdot\bm{S}_{i+1})^2
\end{equation}
Hamiltonian \eqref{eq:HBB} is derived within second order perturbation theory in the tunneling rate and the angle $\theta\in[-\pi;\pi]$ depends 
only  on the two-body short range atom-atom interaction.
This model has a rich phase diagram depending on the angle $\theta$ and has  been extensively studied in the literature, see \cite{AKLT, FathSolyom1991,dimer, Schollwock1996,Lauchli2006} and references therein. We review the phase diagram, depicted in Fig.~\ref{fig:2}(a):

\emph{The ferromagnetic phase.-} For $\pi/2 <\theta < 5\pi/4$, the ground state is ferromagnetic exhibiting a spontaneous magnetization, which serves as a local order parameter.\\
For the remaining values of $\theta$, the ground state has no net magnetization but different quantum phases:\\
\indent\emph{The critical phase.-} In the interval $\pi/4<\theta<\pi/2$ the system is gapless due to soft collective modes at momenta $q = 0,\pm 2\pi/(3d)$. The ground state spin-spin correlation functions $\langle S_{zi}S_{z(i+r)}\rangle $ show period-$3$ oscillations \cite{FathSolyom1991}. In momentum space, this feature emerges as a peak at $q= 2\pi/(3d)$ in the magnetic structure factor $S(q)$. 
Recently L\"auchli et al. \cite{Lauchli2006} have shown that nematic (i.e. quadrupolar) correlations at momentum $q= 2\pi/(3d)$ are enhanced in the critical phase while spin correlations become smaller when increasing $\theta$ from $0.2\pi$ to $0.5 \pi$.  Together with the absence of the gap, the enhanced nematic correlations are a distinctive feature of the critical phase. 
 
\emph{The Haldane phase.-}  The interval $-\pi/4<\theta<\pi/4$ contains for $\theta=0$ the spin-1 Heisenberg chain and for $\tan(\theta) = 1/3$ the exactly solvable AKLT point \cite{AKLT}.  The presence of a gap excludes local quasi long range  order and spin correlations decay exponentially. The Haldane phase can be characterized by a hidden topological order parameter, called the string order parameter \cite{Rommelse}, that cannot be revealed with local measurements.

\emph{The dimer phase.-}   In the region $-3\pi/4<\theta<-\pi/4$ the ground state is gapped and breaks translational invariance, organizing in slightly correlated dimers.
For $-3\pi/4<\theta<-\pi/2$ it is still under debate whether the system is always dimerized or it becomes  nematic \cite{nematic}. Numerical results \cite{dimer,Lauchli2006} show that the dimer order parameter,  $D= |\langle H_i-H_{i+1}\rangle|$ where $H_i=\cos(\theta) \bm{S}_i\cdot\bm{S}_{i+1}+ \sin(\theta) (\bm{S}_i\cdot\bm{S}_{i+1})^2$, is different from zero up to values very close to $\theta=-3\pi/4$.

\begin{figure}[t!]
\begin{center}
\includegraphics[scale=0.73]{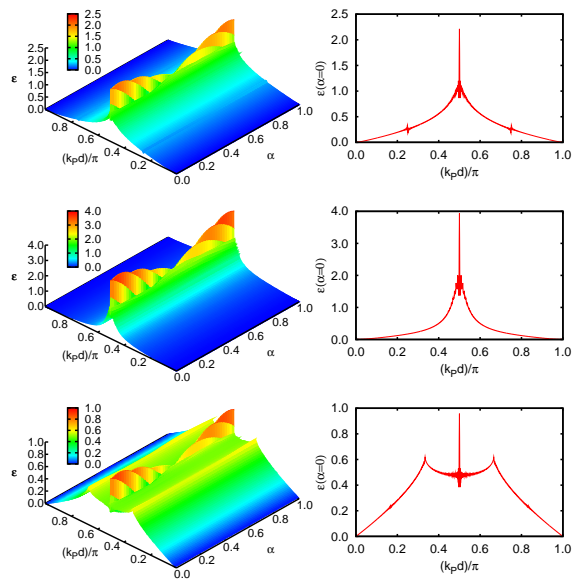}
\caption{(Color online) Left column, the function $\varepsilon(k_P,\alpha)$ (in units of $\kappa^2$) for different values of $\theta$ in the three phases for $L=132$: top $\theta = -0.5\pi$ (dimer), middle $\theta=0$ (Haldane), bottom $\theta=0.3\pi$ (critical). Right column, the same plots but restricted to $\alpha=0$.}
\label{fig:epsall}
\end{center}
\end{figure}

Here we calculate the variance $\varepsilon(k_P,\alpha)$, which depends on all possible two-spin correlations for the ground state of the system at different points of the phase diagram given by $\theta$. Ground states are found numerically using a density matrix renormalization group (see for example \cite{dmrg}) with open boundary conditions. In Fig.~\ref{fig:epsall} we show  $\varepsilon(k_P,\alpha)$ for three different values of $\theta$ corresponding to the critical, Haldane and dimer phases. A common feature of the three phases is the presence of a high peak at $k_P d=\pi/2$ due to antiferromagnetic correlations. Such peak is instead absent in the ferromagnetic phase. Apart from this, the plots in the three phases are qualitatively different. In fact, in the critical phase, the signal is characterized by peaks at $k_Pd\sim\pi/3$ and $k_Pd\sim2\pi/3$. These resemble the peaks of the magnetic structure factor $S(2k_P)$ \cite{Schollwock1996} and are due to the period-3 oscillations of the correlation functions. These correlations are fundamental for the detection of the critical phase. For $\theta < -\pi/4$ we find the appearance of other small peaks at $k_P d = \pi/4$ and $k_P d = 3\pi/4$ signaling, as we show below, the dimer order. Finally, in the Haldane phase, the signal $\varepsilon(k_P,\alpha)$ lacks of non trivial features for $k_P d\neq\pi/2$.

We now use the quantity $\Delta\varepsilon(k_P,\alpha_1,\alpha_2)$ introduced in Eq.~\eqref{eq:deltaeps} for the detection of the critical and dimer phase.
To see how to choose the parameters $k_P,\alpha_1,\alpha_2$, let us consider the critical phase. In this case it is natural to take $k_P=\pi/3d$. Then we study the behavior of $\varepsilon(\pi/3d,\alpha)$ in the critical phase as a function of $\alpha$.

\begin{figure}[t]
\begin{center}
\includegraphics[scale=0.34]{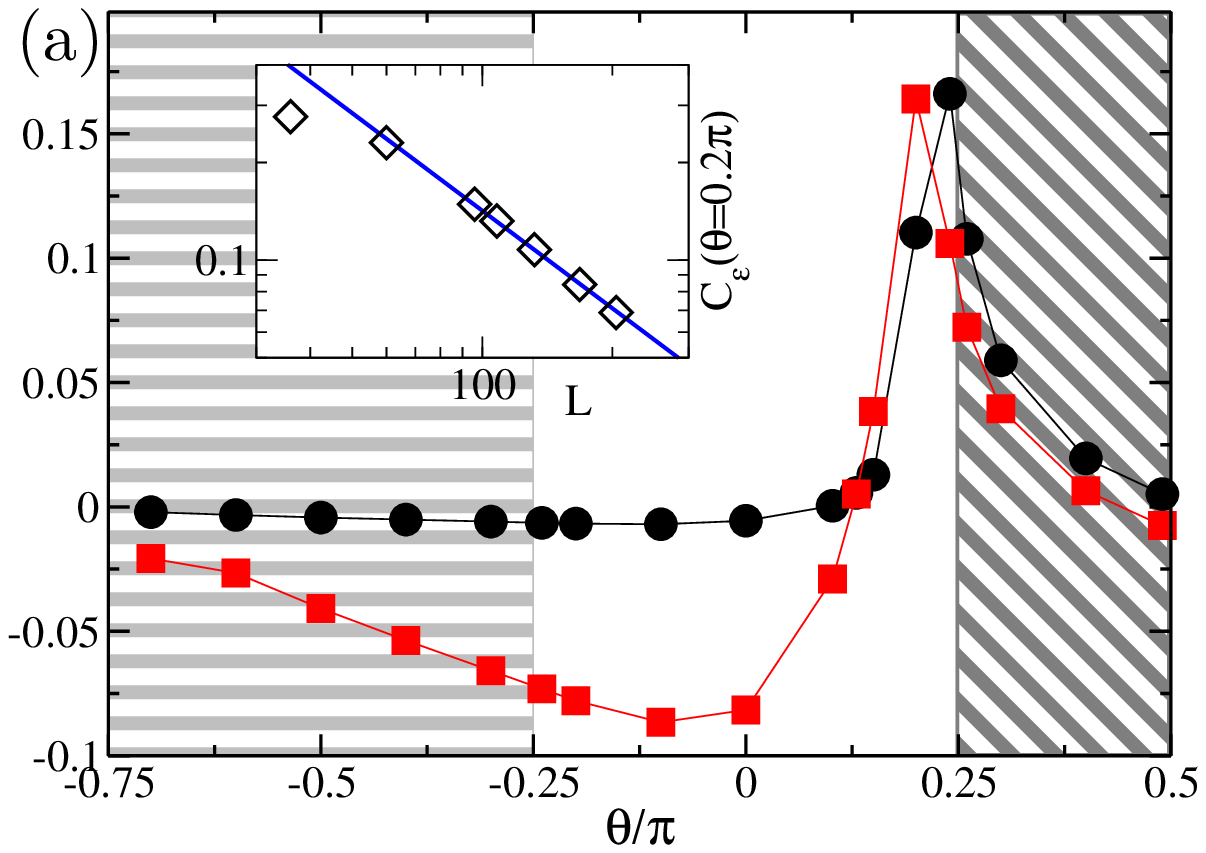}
\includegraphics[scale=0.34]{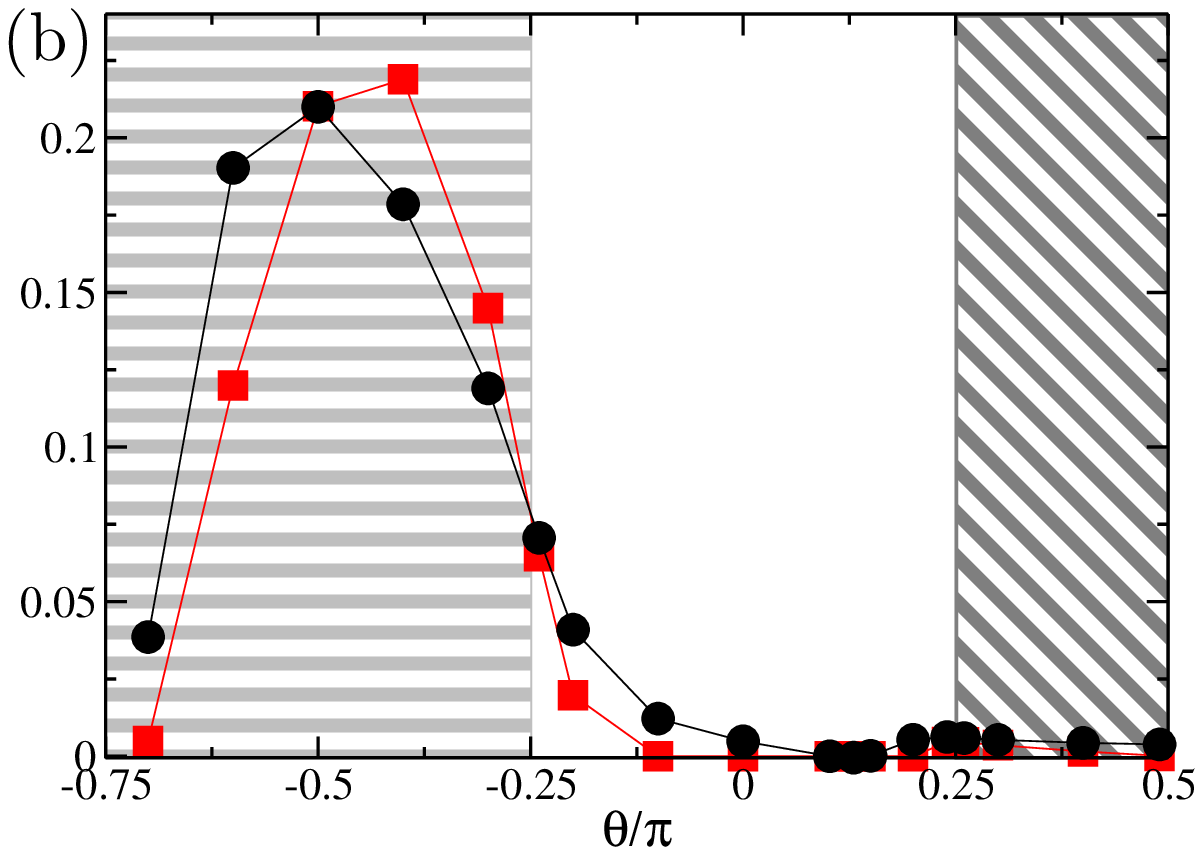}
\caption{(Color online) (a): The quantity $\mathcal C_\varepsilon=\Delta\varepsilon(\pi/3d,5/4,1/2)$, in units of $\kappa^2$, (circles) as a function of $\theta$  for $L=132$ and the rescaled magnetic structure factor $[S(2\pi/3d)-1]/5$ (squares).  Inset: finite size scaling of $\mathcal C_\varepsilon$ for $\theta=0.2\pi$. The solid line scales as $1/L$. (b): the quantity $\mathcal D_\varepsilon = \Delta\varepsilon(\pi/4d,1/2,3/2)$, in units of $\kappa^2$, (circles) for different values of $\theta$ for $L=132$ and the rescaled dimer order parameter $D_R=D/5.38$. In the two plots, we distinguish the model phases with different shading: horizontal lines (dimer), no shading (Haldane), oblique lines (critical).}
\label{fig:results}
\end{center}
\end{figure}

The numerical analysis, reported in Fig.~\ref{fig:2}(b), shows that the quantity $\varepsilon(\pi/3d,\alpha)$ is a sinusoidal oscillating function of $\alpha$. 
The role of correlations at  $k_P=\pi/3d$ is optimized by taking the difference between its maximum value at $\alpha_1=5/4$ and its minimum at $\alpha_2=1/2$. 
We then introduce $\mathcal C_\varepsilon=\Delta\varepsilon(\pi/3d,5/4,1/2)$ which we consider as an order parameter for the critical phase. Since the ground state has zero angular momentum, we obtain:
\begin{eqnarray}
\mathcal C_\varepsilon&=&
\frac{1}{L} \sum_{mn} \cos\left[\frac{2\pi}{3} (m+n) +\frac{\pi}{3}\right] \mathcal G_z(m,n)
\end{eqnarray}
The quantity $\mathcal C_\varepsilon$ is sensitive to correlations which oscillate with a period 3 and represents a footprint of the critical phase. In Fig.~\ref{fig:results}(a) we show the signal $\mathcal C_\varepsilon$ for different values of $\theta$. The results clearly show that the critical phase is very well detected by a positive value of  $\mathcal C_\varepsilon$. For $\theta=0.2 \pi$, in the Haldane phase and close to the phase transition, we still observe a large positive value, probably due to residual period-3 correlations persisting in the Haldane phase. However, in this point, we find a non negligible dependence with the size of the sample. As shown in the inset of Fig.~\ref{fig:results}(a) a finite size scaling suggests that in the thermodynamical limit for $L\to\infty$ the quantity $\mathcal C_\varepsilon$ goes to zero as $1/L$ for $\theta=0.2\pi$, while for the other values of $\theta \ge 0.24 \pi$ it converges to a finite value.
In Fig.~\ref{fig:results}(a) we compare $\mathcal C_\varepsilon$ with the magnetic structure factor $S(2\pi/3d)$. We confirm the results found in \cite{Lauchli2006} for chains up to $L=18$: since the maximum is at $\theta\simeq 0.2\pi$, the magnetic structure factor alone is not sufficient to distinguish the critical point at $\theta=0.25\pi$. As shown in \cite{Lauchli2006}, one would also need the quadrupolar structure factor to find the critical point. Our findings indicate that measuring only $\mathcal C_\varepsilon$, which behaves as an order parameter, is sufficient to infer the occurrence of the phase transition. 

In the dimerized phase, the presence of peaks at $k_Pd=\pi/4$  signals the pairing of neighboring spins. By averaging the signal $\varepsilon(k_P,\alpha)$ over $\alpha$, these peaks disappear. Therefore, these features are not visible in the magnetic structure factor. Following a similar reasoning that leads to the definition of $\mathcal C_\varepsilon$, we find that the quantity 
$$
\mathcal D_\varepsilon \equiv\Delta\varepsilon(\frac{\pi}{4d},\frac 12,\frac 32)
= -\frac 1L \sum_{mn} \sin\left[\frac{\pi}{2} (m+n)\right] \mathcal G_z(m,n)
$$
 is suitable for the detection of the dimer phase. The factor $\sin\left[\pi (m+n)/2\right]$ ensures that only the pairs of spins with positions $m$ and $n$ of opposite parity contribute to $\mathcal D_\varepsilon$.  The differential signal $\mathcal D_\varepsilon$ is therefore an extension to long range correlations of the dimer order parameter $D$.
In Fig.~\ref{fig:results}(b) we compare the signal $\mathcal D_\varepsilon$ and the dimer order parameter $D$ for different values of $\theta$. Similar to $D$, the quantity $\mathcal D_\varepsilon$ is significantly different from zero only in the dimerized phase, with the advantage of being experimentally measurable.

The simulation of strongly correlated magnets with optical lattices can be foreseen in the next generation of experiments with ultracold atoms. Once this is accomplished, the measurement of order parameters characterizing these phases will be of utmost importance.
In this work we proposed the use of QPS for the direct observation of magnetic order parameters in  ultracold atoms in optical lattices. As a proof of principle, we have shown that this method allows us to unambiguously reconstruct the rotational invariant spin-1 chain phase diagram. We emphasize that the method is very general and could be applied to other spin models. Notice also that the QPS technique we propose is not based on any previous knowledge of the quantum phase to be characterized, but rather takes advantage of the features displayed by the probing scheme properly optimized. Although we are not able to measure directly the string order parameter in the Haldane phase with second order correlation functions, topological order can in principle be extracted from the higher moments of the output light quadrature. This measure, provided that the shot noise of the incoming light is small, employing for instance squeezed light, adds no experimental effort and will be the subject of future investigations.\\
{\it Acknowledgments.--} We thank P. Massignan, S. Paganelli, M. Rizzi, and U. Schollw\"ock for discussions. We acknowledge support from the Spanish MICINN (Juan de la Cierva, FIS2008-01236 and QOIT-Consolider Ingenio 2010), Generalitat de Catalunya Grant No. 2005SGR-00343 and the Alexander von Humboldt foundation. We used the DMRG code available at \url{http://www.dmrg.it}.


\begin{thebibliography}{99}
\bibitem{BEC}
M. H. Anderson et al., Science {\bf 269}, 198 (1995); K. B. Davis et al., Phys. Rev. Lett. {\bf 75}, 3969 (1995).

\bibitem{reviewAnna}
M. Lewenstein et al.,
Adv. in Phys. {\bf 56},243 (2007). 


\bibitem{reviewBloch}
I. Bloch, J. Dalibard, and W. Zwerger, Rev. Mod. Phys. {\bf 80}, 885 (2008).

\bibitem{TrotzkyAndKetterle}
S. Trotzky et al., Science {\bf 319}, 295 (2008).
P. Medley et al., arxiv:1006.4674



\bibitem{Altman2004}
E. Altman, E. Demler, and M. D. Lukin, Phys. Rev. A {\bf 70}, 013603 (2004).

\bibitem{Bloch2005}
S. F\"olling, et al.,
Nature {\bf 434}, 481 (2005).



\bibitem{Hulet2010}
T. A. Corcovilos, et al.,
Phys. Rev. A  {\bf 81}, 013415 (2010).

\bibitem{Burnett}
J. Douglas and K. Burnett, arxiv:1007.1899.

\bibitem{markusandbloch}
W. S. Bakr et al., 
Nature {\bf 462} 74 (2009);
J. Sherson et al., 
 Nature {\bf 467}, 68 (2010).



\bibitem{Eckert2008}
K. Eckert, et al.,
Nat. Phys. {\bf 4}, 50 (2008). 

\bibitem{Julsgaard}
B. Julsgaard,
PhD-thesis, University of Aarhus (2003).


\bibitem{Roscilde2009}
T. Roscilde, et al.,
New. J. Phys. {\bf 11}, 055041 (2009).

\bibitem{Imambekov}
A. Imambekov, M. Lukin, and E. Demler, 
Phys. Rev. A {\bf 68}, 063602 (2003).


\bibitem{dimer}
G. F\'ath and J. S\'olyom, Phys. Rev. B {\bf 51}, 3620 (1995).
M . Rizzi, et al.
Phys. Rev. Lett. {\bf 95}, 240404 (2005).
K. Buchta et al.,
Phys. Rev. B {\bf 72}, 054433 (2005).

\bibitem{Lauchli2006}
A. L\"auchli, G. Schmid, and S. Trebst,
Phys. Rev. B {\bf 74}, 144426 (2006).

\bibitem{AKLT}
I. Affleck, et al.,
Phys. Rev. Lett. {\bf 59}, 799 (1987).


\bibitem{Schollwock1996}
U. Schollw\"ock, Th. Jolicoeur, and T. Garel,
Phys. Rev. B {\bf 53}, 3304 (1996).

\bibitem{FathSolyom1991}
G. F\'ath and J. S\'olyom, Phys. Rev. B {\bf 44}, 11836 (1991).


\bibitem{Rommelse}
M. den Nijs and K. Rommelse,
Phys. Rev. B {\bf 40}, 4709 (1989).

\bibitem{nematic}
A. V. Chubukov, Phys. Rev. B {\bf 43}, 3337 (1991); 
B. A. Ivanov, A. K. Kolezhuk, ibid. {\bf 68}, 052401 (2003); 
T. Grover and T. Senthil, Phys. Rev. Lett. {\bf 98}, 247202 (2007);




\bibitem{dmrg}
U. Schollw\"ock, Rev. Mod. Phys. {\bf 77}, 259 (2005);
G. De Chiara, et al.
J. Comp. Theor. Nanos. {\bf 5}, 1277 (2008).
\end{thebibliography}
\end{document}